# THE OPTICAL INTERFACE OF A PHOTONIC CRYSTAL: MODELING AN OPAL WITH A STRATIFIED EFFECTIVE INDEX

**Isabelle MAURIN**[1,2], **Elias MOUFAREJ**[1,2], **Athanasios LALIOTIS**[1,2], **Daniel BLOCH**[2,1]

(1) Laboratoire de Physique des Lasers, Université Paris 13, Sorbonne Paris-Cité, F-93430 Villetaneuse, France

(2) CNRS, UMR 7538, 99 Avenue J-.B. Clément, F-93430 Villetaneuse, France

e-mail: isabelle.maurin@univ-paris13.fr

*ABSTRACT:*

*An artificial opal is a compact arrangement of transparent spheres, and is an archetype of a three-dimensional photonic crystal. Here, we describe the optics of an opal using a flexible model based upon a stratified medium whose (effective) index is governed by the opal density in a small planar slice of the opal. We take into account the effect of the substrate and assume a well- controlled number of layers, as it occurs for an opal fabricated by Langmuir-Blodgett deposition. The calculations are performed with transfer matrices, and an absorptive component in the effective index is introduced to account for the light scattering. This one-dimensional formalism allows quantitative predictions for reflection and transmission, notably as a function of the ratio between the irradiation wavelength and the sphere diameter, or as a function of the incidence angle or of the polarization. It can be used for an irradiation from the substrate side or from the vacuum side and can account for defect layers. The interface region between the opal and the substrate (or vacuum) is shown to have a strong influence, regardless of the exact opal structure. This break in the periodicity at the*

*interface is a general, but often ignored feature, of any external coupling to a photonic crystal. Our calculations provide also the main features of the Bragg peak for reflection, including its width and strength. Comparisons of this versatile model with experiments show that despite its simplicity, it is powerful enough to explain numerous observations.*

I.  **INTRODUCTION**

Structures with a refractive index periodically varying on a scale comparable to an optical wavelength are described as "photonic crystals" [1], and photonic energy bands (allowing or blocking light transmission) are predicted. The interest in photonic crystals is usually connected to their "bulk" optical properties, with the expectation of specific propagation rules *inside* the periodic arrangement of the crystal and related prohibited emission. Actually, the optical information that can be obtained from a photonic crystal requires a detection *outside* the crystal. It originates either from an internal emission which is transmitted or scattered outside, or from an external irradiation which has entered into the crystal, or has been reflected by it. Hence, the interface with the environment can critically affect the detection of the optical properties of the photonic crystal: (i) as a principle, this boundary region tends to break the periodicity typical of the photonic crystal; (ii) optical reflection at the interface, which cannot be ignored when the detector is external to the crystal, usually probes the optical response on a dimension related to the optical wavelength $\lambda$. For these reasons, experimental tests of the quality of a photonic crystal are at a risk to explore mostly the quality of the first layers only, notably when the tests are conducted in reflection.

Many techniques have been developed to solve the difficulty of fabricating three-dimensional or two-dimensional photonic crystals [2]. Soft chemistry methods [3-8] are convenient in spite of a limited choice of geometry and materials. In particular, artificial opals can be obtained through the self-organization of sub-micrometric mono-disperse spheres (in polystyrene, silica, or even $TiO_2$) in a process of sedimentation or of convection, implying the evaporation of a solvent [3-6] and deposition of the spheres onto a substrate. A microscopic description of the evaporation of the solvent justifies that artificial opals, produced as a bulk material, tend to organize as a compact face-centered cubic (*f.c.c.*) crystalline arrangement [6] of dielectric spheres. The three-dimensional organization can be observed for good quality opals, with specific signatures associated to the organization of a (111) or (200) crystalline plane [9-11]. A possible alternate fabrication technique is the Langmuir-Blodgett (LB) deposition, which consists of a successive layer-by-layer deposition and allows a control of the number of deposited layers [7, 8]. In such a technique, the photonic crystal, always sensitive to the dispersion in the sphere size, is usually highly polycrystalline, notably in the direction normal to the LB deposition plane, and the crystalline arrangement usually exhibits a random hexagonal close-packed (*r.h.c.p.*) structure (*i.e.* a mix of *f.c.c.* and *h.c.p.* hexagonal close-packed lattices rather than a true *f.c.c.* arrangement). In all cases, the opal is not self-supported, but has grown from a (planar) substrate, with the (111) plane parallel to the substrate. It is an aim of the present work to propose a simple model for the optics of an artificial opal, which has intrinsically to be deposited on a substrate.

In a previous work [12], performed in the context of infiltrating a resonant vapor in the interstitial regions of an opal, we have already performed optical experiments on a glass opal deposited on a substrate. Typically, we irradiated an opal made of 10 or 20 layers of spheres whose diameter ($D = 1$ μm, or $D = 400$ nm) compares with the wavelength of the light ($\lambda = 852$ nm, 894 nm, or even 455 nm). In addition to the scattering of the incident light - whose residual coherence occasionally leads to specific diffraction figures, such as an hexagonal diffraction for a single opal layer [13], or to more complex structures for a perfect arrangement -, we had noted that a fraction of the beam power undergoes a specular reflection at the substrate/opal interface, while another part of the beam is transmitted. The presence of such beams, although well documented [4, 9-11, 14-17] is partly unexpected because the opal/substrate and opal/vacuum interfaces are strongly non planar (at a microscopic/wavelength scale). Also, the incident beam polarization (assuming a principal polarization for the irradiation) is mostly conserved [18] for these reflected or transmitted beams [19], and we had observed [12] that the quantitative evolution of their intensities with the incidence angle exhibits trends similar to those predicted by Fresnel formulae, including a nearly null reflection analogous to a Brewster incidence.

To understand the optics of an opal, or more generally of a photonic crystal deposited on a substrate, a bandgap calculation solely based upon the opal periodicity (and indices, sphere diameter...) would not be sufficient because of the interfaces problem, and the related break in periodicity. Rather, *ad hoc* models are developed in the literature, based upon numerical calculations such as finite element methods (see [20]). An inconvenience of these heavy methods is their very limited versatility and weak robustness: the effect of a small change in the parameters of the opal can hardly be predicted from a linear variation. This may even mean that the comparison between predictions and experiments can be impaired by uncertainties, such as the ones concerning the sphere diameter, or the actual index of the glass spheres as they currently exhibit some porosity. Moreover, these crystal-based models have difficulties [21] incorporating the random defects and imperfections of a self-organized material, which affect the periodicity of the crystalline arrangement of successive layers and the sphericity of glass balls, leading to a poly-domain crystal. Despite these limitations, an essential benefit of three-dimensional models is that they have the ability to predict the light reflectivity and transmission, and the scattering as well, with its spatial distribution of intensity, *i.e.* speckle and multiple diffraction as due to coherence. Alternately, simplified models dealing with a homogeneous "effective index" have been popular in the photonic crystals literature [22] and about opals since [4], and can be considered a "zero-dimensional" model of the crystal periodicity [10, 17]. In this approach [4], the concept of "effective index", developed in optics to take into account inclusion of "small-size" crystallites (*i.e.* small relatively to the relevant optical wavelength), is actually extended to the glass spheres (or alternately to the vacuum interstices between spheres of the opal), whose size is of the same order of magnitude as the wavelength. Such an approach has even been proposed to describe antireflection properties of a single layer opal [23, 24], with the inconvenience that the thickness is arbitrarily taken as the height of the single layer of opal, and that no parameter adjusts the variations with the incidence. First introduced in [4], it is also common to find models combining the geometrical periodicity of the opal crystalline arrangement, responsible for a peak of reflectivity analogous to a Bragg diffraction peak, with the suitable "effective index" model to calculate the optical periodicity inside the opal [10, 17]. Adjustments between the model and experiments performed under various incidences are sometimes used to produce a refined estimate of the sphere diameter, and/or of the sphere index as indication of its porosity. Apart from the correlation between the "Bragg angle" and the irradiation wavelength, such an approach is unable to yield quantitative predictions regarding the transmission, reflection or scattering.

In the present work, we consider a stratified one-dimensional version of an effective index model. This allows a much better description of the interfacial regions between the opal and the substrate (or the vacuum), notably for these extreme regions where the glass spheres are not in a compact arrangement. Such a model, although mentioned as a principle in literature [19], has not been fully developed until now [25]. Among several advantages, including rather light and versatile one-dimensional calculations, it appears well-suited to a thin LB opal with its already mentioned *r.h.c.p.* structure. Naturally, the ambition of such a one-dimensional index model is limited to a fair description of reflection and transmission behaviours, and the light scattering and diffraction cannot be evaluated. Rather, an *ad hoc* loss must be added to the model to avoid the sum of the reflection and transmission coefficients to be unity for a transparent material like glass. With our model, we are able to discuss the physical features associated to the interface with the opal (notably the influence of the "gap" region between the substrate or vacuum, and the opal), the build-up of a quasi "Bragg diffraction" and the effect of an imperfect opal periodicity. An extension of these results to the properties of a resonant material infiltrated in the void regions of an opal [26] can be considered, and is the topic of a further paper [27]. Results from our previous experiments [12], and from complimentary dedicated experiments, are shown to be compatible with our quantitative predictions once the extinction parameter is reasonably adjusted.

The paper is organized in the following way: in section II, we describe the opal deposited on a substrate as a stratified medium, and introduce a formal optical treatment model through transfer matrices, providing the reflection and transmission coefficients. The extension to complex stratified indices allows replacing scattering by an absorption coefficient. Practically, reflection and transmission by the opal are mostly governed by parameters such as the ratio $\lambda/D$ between the irradiation wavelength and the sphere diameter. On this basis, and considering several types of distribution of the stratified index, we discriminate in section III the contribution of the different regions, notably discussing the specific influence on reflection of the interface between substrate (or vacuum) and the opal, the coherent construction of a Bragg reflection associated to the (bulk) periodicity of the opal, and the effect of a single layer defect introduced on purpose or as a fabrication defect. The next section (section IV) reports on complimentary experiments measuring the reflection and transmission spectra of opals of a smaller size, so that the Bragg reflection is observed and compared to the predictions of the model. Finally, in the conclusive section (section V), we summarize the results and consider some possible extensions such as inverse opals and photonic crystals based on cylindrical rods.

## II) OPTICAL MODEL OF THE OPAL AND FORMALISM

### 1) Stratified effective index

Effective index model is currently applied to average, through a single parameter, the optical properties of a medium that includes voids or impurities, whose size is much smaller than the optical wavelength ($\lambda$). The effective index $n_{eff}$ is deduced from an "averaged" permittivity. For a dielectric medium with small size (vacuum) voids, it is hence defined as:

$$n_{eff} = \sqrt{f\varepsilon + (1-f)} \qquad (1)$$

with $\varepsilon$ the dielectric constant of the filling material ($\varepsilon = n^2$ with $n$ the dielectric index of the filling material), and $f$ the filling factor of the dielectric medium.

To extend an "effective index" description to a "Stratified effective index" model for the opal, one has to "slice" the opal in successive parallel layers, distributed along the direction $z$ perpendicular to the substrate. This requires to consider a spatially-dependent (along $z$) filling factor $f(z)$ of the dielectric material (spheres for opal), and the stratified effective index distribution is hence given by $n_{eff}(z)$:

$$n_{eff}(z) = \sqrt{f(z).\varepsilon + (1 - f(z))} \qquad (2)$$

### 2) **Sphere packing in an opal and periodicity**

An ideal opal is a close-packed arrangement of identical spheres, which can be described as successive layers of bi-dimensional sections of close packed spheres (a single layer of spheres). The successive layers themselves are arranged in a compact manner, so that the distance between two successive layers is $D\sqrt{\frac{2}{3}}$ [25] with $D$ the sphere diameter, although the LB opal is not a three-dimensional crystal (*i.e.*, with its *r.h.c.p.* structure, the LB opal is not regularly organized along $z$).

For the first layer above the substrate (which we locate in the $z \leq 0$ region), the filling factor, which corresponds to a closed-packed distribution of circles in hexagonal cells in the equatorial plane ($z = D/2$), is given by:

$$f_1(z) = \frac{2\pi z(D-z)}{\sqrt{3}D^2} H(z) \qquad (3)$$

with $H(z) = 1$ for $D \geq z \geq 0$ and $H(z) = 0$ elsewhere

For an opal made of *N* layers arranged in a compact manner, the filling factor *f(z)* is the sum of the filling factor of the individual layers.

$$f(z) = \sum_{i=1}^{N} f_1\left[z - (i-1)D\sqrt{\tfrac{2}{3}}\right] \qquad (4)$$

In Fig. 1, we show the filling factor for an opal of glass spheres (fig 1a), and the corresponding effective index (fig 1b). Both structures are very similar: one recognizes a $D\sqrt{2/3}$ periodicity between the first equatorial plane (at $D/2$) and the last one. The structure of a periodic layer exhibits a maximal value in the equatorial planes of spheres, and periodic relative extremes associated to the compact packing between successive layers, when two terms in eq.4, and not a single one, are non-zero.

Also, Fig.1 clearly shows that the "opal and substrate system" combines the periodicity of the equatorial planes ($D\sqrt{2/3}$ because we have assumed a compact opal) and nearly empty regions (of a thickness $D/2$) when there is only a point contact between the spheres and the substrate. The consequences of this periodicity break, as induced by this interface region for the first (and last) half-layer, will be discussed in detail in section III with the help of the formalism that we develop below.

Note that a layered calculation usually considers layers of finite thickness (fig. 1b), so that the continuous functions $f(z)$, and $n_{eff}(z)$, determined here for an ideal opal, have to be discretized in sufficiently small steps. Also, the layered effective index formalism is not limited to a particular shape as defined for *f(z)* in eq. (4), but may as well account for a non-compact arrangement as well as for crystal defects.

## 3) Propagation in a stratified medium

a) <u>Matrix formalism and reflection/transmission coefficients</u>

We recall here the standard matrix formalism for a stratified medium made of successive parallel layers [28, 29], and use it to calculate the reflection and transmission coefficients. We consider an incident light irradiation defined by its electric field:

$$\vec{E} = E_0 . \exp j(\omega t - \vec{k}.\vec{r})\vec{u} \qquad (5)$$

with a (circular) frequency $\omega$ (corresponding to a wavelength in vacuum $\lambda$) in the $z \leq 0$ region of index $n_0$, incident under an incidence $\theta_0$ on a medium composed of N finite parallel layers (perpendicular to $z$, in the $z \geq 0$ region), and ended by a region of index $n_{N+1}$. In eq. (5), $\vec{k}$ is the wave vector and $\vec{u}$ the unit vector giving the direction of the polarization.

The matrix formalism allows going from one layer to a neighboring one with the limiting boundary conditions, namely the continuity of the tangential components of the total electromagnetic field (*i.e.* electric and magnetic fields $E$ and $H$) at an interface. It takes into account the propagation in each layer, for the forward and backward field components resulting from successive transmission and reflection and which propagate under an angle satisfying the Snell's law at the successive interfaces. The two principal modes of polarization TE and TM, have to be dealt with separately, and an arbitrary input polarization has to be processed as a linear combination of principal polarizations.

Following the notations of Fig. 2, one defines the tangential component of the amplitude component of the electric (respectively magnetic) field at the generic boundary between the $(i\text{-}1)^{th}$ and $i^{th}$ layers as $E_{i-1,i}$ (respectively $H_{i-1,i}$). For the $i^{th}$ layer, the index is defined as $n_i$, the thickness as $d_i$, and the propagation angle as $\theta_i$ with

$$n_0 \sin \theta_0 = n_i \sin(\theta_i) \qquad (6)$$

The transfer matrix $M_i$ then appears when comparing the $(i\text{-}1)^{th}$ and $i^{th}$ layers boundary, with the $i^{th}$ and $(i+1)^{th}$ layers boundary. One finds:

$$\begin{bmatrix} E_{i-1,i} \\ H_{i-1,i} \end{bmatrix} = \begin{bmatrix} \cos(\delta_i) & j\sin(\delta_i)/Y_i \\ jY_i \sin(\delta_i) & \cos(\delta_i) \end{bmatrix} \begin{bmatrix} E_{i,i+1} \\ H_{i,i+1} \end{bmatrix} \quad (7)$$

so that

$$M_i = \begin{bmatrix} \cos(\delta_i) & j\sin(\delta_i)/Y_i \\ jY_i \sin(\delta_i) & \cos(\delta_i) \end{bmatrix} \quad (8)$$

In eqs. (7-8), one has defined:

$$\delta_i = \tfrac{2\pi}{\lambda} n_i d_i \cos\theta_i \quad (9)$$

and $Y_i$ depends on the polarization:

for TE polarization: $Y_i = n_i \cos(\theta_i)$  (10)

for TM polarization: $Y_i = n_i / \cos(\theta_i)$  (11)

For N layers, the total matrix $M$ is the product of individual transfer matrices $M_i$, so that the input and output tangential fields can be calculated by multiplying the different transfer matrices together:

$$\begin{bmatrix} E_{0,1} \\ H_{0,1} \end{bmatrix} = M \begin{bmatrix} E_{N,N+1} \\ H_{N,N+1} \end{bmatrix} \quad (12)$$

with

$$M = \begin{pmatrix} A & B \\ C & D \end{pmatrix} = M_1 M_2 ... M_N \quad (13)$$

The tangential components of the field at the input and output boundaries can be rewritten:

$E_{0,1} = E_0 (1 + r)$  (14)

$E_{N, N+1} = t \cdot E_0$  (15)

$H_{0,1} = E_0 (1 - r) Y_0$  (16)

$H_{N,N+1} = E_0 \, t \, Y_{N+1}$  (17)

with $r$ and $t$ the standard reflection and transmitted amplitude coefficients $r = \dfrac{E_r}{E_0}$ and $t = \dfrac{E_t}{E_0}$,

with $E_r$ and $E_t$ defined in fig. 2

One hence deduces:

$$r = \frac{Y_0 A + Y_0 Y_{N+1} B - C - Y_{N+1} D}{Y_0 A + Y_0 Y_{N+1} B + C + Y_{N+1} D} \quad (18a)$$

$$t = \frac{2Y_0}{Y_0 A + Y_0 Y_{N+1} B + C + Y_{N+1} D} \quad (18b)$$

from which the reflection and transmission intensity coefficients are simply calculated:

$$R = r.r^* \quad (19)$$

$$T = \frac{n_{N+1} \cos(\theta_{N+1})}{n_0 \cos(\theta_0)} .tt^* \quad (20)$$

b) <u>Input and output media for an opal deposited on a substrate</u>

The problem that we address is the one of an opal deposited on a transparent substrate, like a glass (parallel) window. Two different situations are of interest: light can enter from the substrate side (as in our experimental work [12] with a resonant gas), or from vacuum (or air). In all cases, the medium outside the opal, considered to be infinite in the matrix formalism described above, is transparent; also, the incidence of the input beam governs all beam directions (see eq. 6). For convenience, and to help comparing the cases when light enters from the substrate, or from the vacuum region, we will always use in the following the "external" incidence angle. The "external angle" $\theta$ is the angle in vacuum, and it is assumed that if the substrate is on the left side (medium described by a $n_0$ index) in Fig.2, there is actually, somewhere in the z < 0 half-space, a vacuum region for which the light enters into the window. This indeed corresponds to the situation of a real experiment (*i.e.* a substrate with a finite width, but not parallel enough to generate interferences). The "external angle" is hence defined by:

$$\theta = \sin^{-1}[n_0 \sin \theta_0] \quad (21)$$

A consequence of our assumption that the substrate itself is actually not infinite, but is a nearly parallel slab, is that this implies that $n_0 \sin \theta_0 \leq 1$. Note that an interesting extension, allowing a larger range of effective incidence angles (*i.e.* not restricted to $\sin \theta_0 \leq 1/n_0$) could be considered for an opal that would be deposited on a prism-shaped substrate.

c)   Extension to absorptive layers

As already mentioned in section I, if the stratified media are assumed to be transparent, one should get a R + T = 1 conservation law, which is unacceptable because of scattering. This is why we introduce phenomenological losses (*i.e.* an absorption coefficient $\alpha_i$) for the $i^{th}$ layer) in the stratified index model, through a complex index $N_i$:

$$N_i = n_i - j\kappa_i = n_i - j\frac{\alpha_i \lambda}{4\pi} \quad (22)$$

Note that it is only when the absorption integrated over a single layer remains small that the stratified model can be useful to describe the opal structure: indeed, a too strong scattering on a single layer of spheres would make it very difficult to recognize an effect of the crystalline organization along *z* of the opal.

The formalism of the sub-section above still applies with complex index stratified media. The real angle $\theta_i$ is replaced by a complex value, that we note $\theta_i^{'}$ and which no longer represents a direction of propagation [29, 30]. However, $\theta_i^{'}$ still follows the Snell's law (see eq. (6)), and significantly, the quantity $N_i \sin \theta_i^{'}$ ($= n_0 \sin \theta_0$) remains real because the input and output media (vacuum or glass) are transparent ($n_0, n_{N+1}$ are real). We show below that under the assumption of a weak absorption ($\kappa_i \ll 1$), the propagation direction remains unchanged. Indeed, the phase-term $\delta_i$ appearing in the matrices of eqs. 7-8, given without absorption by:

$$\delta_i = \tfrac{2\pi}{\lambda} n_i d_i \cos \theta_i = \tfrac{2\pi}{\lambda} d_i \sqrt{n_i^2 - n_0^2 \sin^2 \theta_0} \quad (23)$$

becomes complex with absorption:

$$\delta_i = \tfrac{2\pi}{\lambda} N_i d_i \cos\theta_i' = \tfrac{2\pi}{\lambda} d_i \sqrt{(n_i - j\kappa_i)^2 - n_0^2 \sin^2\theta_0}) \quad (24)$$

so that with the $\kappa_i \ll 1$ approximation, one has:

$$\delta_i \approx \tfrac{2\pi}{\lambda} d_i \sqrt{n_i^2 - n_0^2 \sin^2\theta_0} - j.\tfrac{2\pi}{\lambda} d_i \frac{\kappa_i n_i}{\sqrt{n_i^2 - n_0^2 \sin^2\theta_0}} \quad (25)$$

which can be rewritten:

$$\delta_i \approx \tfrac{2\pi}{\lambda} d_i n_i \cos\theta_i - j.\tfrac{2\pi}{\lambda} d_i \frac{\kappa_i n_i}{\cos\theta_i} \quad (26)$$

In eq. (26), $\theta_i$ (calculated without absorption) still gives the direction of propagation. It is a remarkable point that while the dephasing associated to propagation decreases as usual with the incidence angle, the field attenuation, provided by $\Im m(\delta_i)$, increases with the incident angle proportionally to the length traveled inside the medium $\frac{d_i}{\cos\theta_i}$.

The phenomenological absorption has been introduced to account for the "scattering losses". It should naturally depend on the wavelength (through the sphere size/ wavelength ratio $D/\lambda$ - or $Dn_{sphere}/\lambda$ inside to consider the wavelength in the medium-), as we will show in section IV. It may depend on the incidence angle (see eq. (26) and below) and on the polarization (TE or TM). Also, the layered model (see eq. 22) can allow for a spatial distribution of the loss $\alpha(z)$ distribution, discretized in the layered approach. When operating our model, we had sometimes used, and compared, a constant loss model [$\alpha(z) = \alpha$], and a model where losses occur only inside the sphere [$\alpha(z) = \alpha f(z)$], or in the voids [$\alpha(z) = \alpha (1 - f(z))$], or where the losses depend on the contact surface [*i.e.* on the sphere perimeter in the considered layer $\alpha(z) = \alpha f(z)^{1/2}$]. As long as the average loss per layer remains unchanged - and small- , the changes induced when varying the model for scattering remain negligible, and it is just simpler to consider a constant loss model.

### III) **MAJOR RESULTS OF THE OPTICAL MODELING**

**1) Typical numerical values and typical behaviour**

To predict the optical behavior of an opal deposited on a substrate, we apply the stratified model developed in section II, with adequate values depending on the experimental conditions, namely polarization (TE, TM, or linear combination), the side of incidence (opal or substrate side), the incidence angle $\theta$ ("external" angle, see subsection II-3 b), and the irradiation wavelength $\lambda$. For the opal, the relevant figures are the number of layers (N) - typically up to 10 or 20, as scattering tends to hinder the role of the deepest layers -, the sphere index $n_{sphere}$ (in all the following, we take numerically $n_{sphere} = 1.4$) from which $n_{eff}(z)$ is deduced, the substrate index $n_{substrate}$, the sphere diameter $D$ -or only the reduced parameter $D/\lambda$ -, and the absorption $\alpha$ - which can depend on $\lambda$, or possibly on $\theta$ -.

For an optimized description of the opal, the numerical calculations, which are performed for a finite number of layers, should converge when increasing the number of slices for a given opal. Practically, it is efficient to divide the opal in a given $N_{step}$ layers - of equal thickness- per period in the periodic region, and in $N_b$ steps to describe the sharper variations of $n(z)$ on the first and ended half layers. For a faster convergence, these latter "slices" in half-layers regions have unequal thicknesses in order to ensure a regular growth for $n(z)$. In these conditions, and when $D$ and $\lambda$ are comparable, we avoid any convergence problems by taking $N_b \sim 50$, and $N_{step} \sim 20\text{-}40$, meaning that a total of less than $10^3$ steps is needed for a rather thick opal made of 20 layers.

To illustrate the major behavior that we will discuss more at length in this section, we first present in figure 3 a typical reflection spectrum, as calculated using the model described in section II. For this typical situation, we have considered an opal made of a large number of layers (N = 20) of glass spheres (with $n_{sphere} = 1.4$ as usual) deposited for simplicity on a similar glass substrate ($n_{substrate}$ = 1.4), and either ignore scattering losses (fig.3a) either choose an absorption independent of $\lambda$, small enough for a single layer, but non negligible for the opal ensemble (fig.3b). Note that figure 3 extends over a large range of $\lambda/D$ values and that for such an extended spectrum, the transparency of the material on such a broad spectrum is highly hypothetical, limiting the practical validity of the present calculation.

Figure 3 permits to distinguish important features. The spectrum is marked by a sharp peak for $\lambda/D$ ~2.06, corresponding to a strong reflection, as if light could not enter the opal. This assertion has obviously to be tempered when absorption is taken into account (fig 3b). This strong reflection is a typical behaviour for a photonic crystal, and will be shown to be a signature of a Bragg diffraction peak, although the opal is not exactly a periodic system. The width of this peak is essentially a consequence of the numerical assumptions in the modelling. At least two other peaks of a smaller amplitude can be identified ($\lambda/D$ ~1.04 and $\lambda/D$ ~0.7). They are associated to high-order Bragg diffraction peaks. Apart from these peaks, reflection remains rather weak and exhibits an oscillating behaviour, which will be shown to originate in interference effects. In the next subsections, we use various types of elementary models of stratified medium, in order to address these major features of Fig.3. This will make us able to discriminate: (i) the effect of the periodicity break in the interface region at the interface by nulling the internal contrast of the periodic regions; (ii) the effect of periodicity (Bragg peak, and Fabry-Perot like effects), by analyzing the effect of changing the finite number of layers, in order to understand how an asymptotic thick opal regime builds up; and (iii) the effect of a local defect of an opal, which is restricted in the framework of a one-dimensional model to a defect layer. For an easy interpretation, the loss coefficient $\alpha$ is chosen to be wavelength-independent in this section.

## 2) **Effect of the interface between the first opal half-layer and substrate or vacuum on reflectivity**

We consider here a specific stratified system, that could be described as a "fused opal", whose index varies according to the local effective index (eqs.2) for a half layer on sphere ($0 \leq z \leq D/2$), and where the periodic region is replaced by a region of constant index $n_0$ (for continuity reasons, we take $n_0 = n(z = D/2) = 1.36$, see Fig 2) on an infinite length (Fig. 4a) - or at least on a length largely exceeding the (scattering) extinction length-. To single out the reflection contribution of the first half-layer, we take an absorption coefficient $\alpha D = 0.1$, whose effect remains weak over the first half-layer, while allowing a finite-size of the "fused opal".

The reflection coefficient is expected to depend on the ratio $\lambda/D$ which governs the ration between the wavelength, and the extension (~ $D/2$) of the "gap" region. Two limits can be predicted : for $\lambda/D \gg 1$, the "gap" region becomes so thin that it can be ignored and the reflectivity can be simply estimated (*e.g.* from Fresnel formulae) from the interface between the substrate – or vacuum - and the constant index of the "opal" with constant index; conversely, for short wavelengths $\lambda/D \ll 1$, light mostly feels the contrast between the substrate and vacuum at the contact plane with the spheres (or a null contrast for an irradiation from the vacuum region), and further accommodates with the slowly varying index. Figures 4 and 5 provide an insight, covering a large range of $\lambda/D$ ratio, of the evolution between those two extreme predicted behaviors, for the two principal polarizations, and for various incidences.

In fig.4 (b, c), reflectivity on the vacuum side in the short wavelength limit is extremely small as expected, and increases in a nearly monotone manner with increasing $\lambda/D$ ratio. For $\lambda/D \rightarrow \infty$, it typically approaches the Fresnel reflection coefficients (which are incidence and polarization dependent) at a vacuum/$n_0$ interface. For reflection on the substrate side (Fig. 4 d, e), we have deliberately chosen a high value $n_{substrate}$ ( = 1.6) to show how one goes from a Fresnel reflection limit $n_{substrate}$/vacuum, to a vacuum/$n_0$ interface when going from the short wavelength limit (*i.e.* long "gap" region) to the long wavelength limit (*i.e.* short "gap" region). The same kind of nearly monotone evolution appears in Fig. 5 showing the influence of the substrate index (calculated for the normal incidence). It is worth noting that in TM polarization (see Fig. 4 c, e), the reflection coefficient remains very small around 50-60° incidence, a behavior reminiscent of a Brewster angle at a single flat interface. In spite of these simple trends, one also observes, when looking more in details on figs 4 and 5, tiny wavy behaviors which have to be attributed to a complex interference effect in a layered medium, even if "smooth" and monotone.

It is worth noting that the behavior for an irradiation on the substrate side (figs. 4 and 5) remains close to the asymptotic regime of small $\lambda/D$ values, where the dominant effect is the "gap" between the substrate and the thin empty region, as long as $\lambda \leq D$. This agrees with our experimental findings, reported in [12] for experiments at $\lambda$ = 852 nm (and $\lambda$ = 894 nm) for opals with $D$ = 1.0 μm deposited on a standard glass, with $n_{substrate} \approx$ 1.5. We had indeed found a reflectivity (~ 4%) extremely close to the one measured on the bare substrate (*i.e.* a region not covered by the opal) under normal incidence. This could be generalized to any incidence angle or polarization as long as the reflection remains rather small (the two reflectivities diverge only above 30-40°, and for TE polarization only). A very small reflection was even observed for large incidence angles and TM polarization, a situation which exhibits analogy with a Brewster incidence angle. In reflection, similar behaviours were found for 10 and a 20 layers opal.

On the opposite, it is only for rather large values of $\lambda/D$, not yet reached for example for the "Bragg reflection peak" evidenced in Fig.3, that the asymptotic small sphere regime ($\lambda/D \gg$ 1) provides a useful indication.

**3) Effects of the periodic region of the opal in the spectra and Bragg reflection**

In Fig. 3, which shows the predictions of our model for a typical situation, we notice the presence of a sharp peak for $\lambda/D \approx 2.06$. It is a well-known feature of an opal that a sharp peak appears in the reflectivity spectrum [4, 7, 10, 14-15], which is associated to a "forbidden band" in a photonic crystal approach, and for which transmission becomes nearly prohibited. This strong reflectivity is often described as a Bragg reflection, in an analogy with the X-ray diffraction on the periodically located point-like nuclei in the structure of an atomic crystal. For an opal, the periodic three-dimensional grating relies on an ensemble of contacted spheres, but the principle of a geometric condition governing the direction of a Bragg diffraction still applies. The standard Bragg condition $k\lambda = 2a \sin(\theta)$, with k an integer, $a$ the distance between successive planes, and $\theta$ the incidence angle, should however include the effect of propagation in the opal as an heterogeneous medium, so that a modified equation [4] is often applied to find the wavelength $\lambda_{max}$ for an opal [9-10, 17]:

$$k \frac{\lambda_{max}}{D} = 2\sqrt{2/3}\sqrt{n_{eff}^2 - \sin^2\theta} \qquad (27)$$

Eq. (27) takes into account the $D\sqrt{2/3}$ geometrical periodicity of the opal, and the Bragg condition is satisfied in a fictitious medium whose index is the averaged effective index $n_{eff}$ (usually estimated with the compact filling factor 74 %), and for which the effective incidence deduced, by the Snell's law, from the external incidence angle $\theta$ (*i.e.* in vacuum). The peak at $\lambda/D \approx 2.06$ in Fig.3 is compatible with a first-order (k=1) Bragg diffraction peak, and the benefit of our model is that our predictions yield an amplitude, width, and lineshape, to this peak.

To better understand the main features predicted for such a diffraction peak, induced by the periodic nature of the opal, notably the successive compact layers, we simplify the description of the opal as a layered medium by replacing the smooth "gap" regions (first and last half-layers) by a single arbitrary step ($N_b$=1). Figure 6, calculated for a square "gap" (see inset) and irradiation on the vacuum side, shows a behavior analogous to the one shown in Fig.3 around the Bragg peak: the amplitudes and widths are comparable. The respective positions obtained using the generalized Bragg law of eq. 27, are marked on Fig.6. They are in very good agreement, with respect to the width, with our predicted reflectivity spectra. The residual lineshape asymmetry around the Bragg peak, which differs from the one appearing in fig.3, is a signature of the complex interferences taking place in the layered medium. It is easily modified with minor changes in the shape of the periodic structure, or keeping a detailed description ($N_b = 50$) of the interface region (see fig. 7 for $N_{layer} = 10$ or for $N_{layer} = 30$).

For a better understanding of the quantitative features (amplitude and width) provided by our model to the reflectivity spectrum around the Bragg peak, we analyze the effect of the number of layers, keeping here a complete description of the interface region. In fig.7, we note that the asymmetry of the spectra is reversed (as compared to Fig. 6), when we include a smooth ($N_b = 50$) description of the interface regions. Also, it is when increasing the number of layers that the contribution of the periodic region becomes predominant. One typically needs at least 3 layers (see Fig.7) to recognize a Bragg diffraction peak in reflection. The peak gets sharper (maximized amplitude and minimized width) when increasing the number of layers, while the position of the peak moves to an asymptotic position, obtained for ~10-20 layers with our realistic choice of parameters. This asymptotic limit of a large number of layers is related to the scattering/extinction parameter.

In addition, a careful look on fig 7 reveals the appearance of a marked small feature for $\lambda/D \approx 1$, for a sufficient number of layers. The shape of this extra structure is analogous to the one of the main peak for $\lambda/D \approx 2.06$, and an evidence of a second order Bragg diffraction.

The corresponding quantitative prediction is however highly sensitive to the wavelength-dependent absorption chosen to describe the scattering.

### 4) **Fabry Perot oscillations and overall thickness of the opal**

Tiny quasi-periodic oscillations appear in figs. 3, 6 and 7. In Fig. 7, their periodicity and amplitude decrease when increasing the number of layers. Quite naturally, they are associated to a Fabry-Perot like effect between the two extreme regions of the opal, as can be quantitatively verified on Fig.7. They tend to cancel for a thick opal because of the extinction by scattering. A similar behavior could be already observed in fig. 3, noting that the absorption taken into account for fig.3b reduces the amplitude of the oscillations on the red side of the Bragg peak. Contrary to the Bragg diffraction peak, these oscillations are typical of a thin opal made of a limited number of layers. These oscillations are often mentioned in the literature [14, 31]. They are sometimes used to evaluate the opal thickness - if unknown-, or to assess the parallelism of the deposited layers.

### 5) **Effects of a defective layer in the opal**

The quality of a real opal arrangement degrades layer after layer due to dispersion in sphere size, clustering of spheres, sphere porosity, arrangement defaults, etc ... To understand the effects of the defects present in opal and in order to approach a realistic situation [21,32], we use the flexibility of our model to introduce an on-purpose defect in the $i^{th}$ layer of the opal, assuming for the index a sudden change to a constant value which locally breaks the periodicity (see the inset of Fig. 8).

Figure 8 illustrates how the shape of the reflectivity spectrum changes with the position of the defect. Whatever the position of the defect layer, the peak reflectivity decreases, and the wavelength of the peak of the Bragg peak is marginally modified. The changes are stronger when the defect is located among the first layers. Indeed, the contribution of the first layers in the spectra is critical for the build-up of the Bragg peak construction, and the absorption (and the physical scattering as well) makes the contribution of the remote layers less important, as already seen in figure 8. Looking in more details, the amplitude, and the frequency of the smaller oscillations, tend to be modified. This is compatible with the idea that these oscillations are due to a Fabry-Perot like effect (see previous subsection)

## IV) **COMPARISON WITH DEDICATED EXPERIMENTS**

We have already mentioned (section III.2) that in the course of our work with a gas-infiltrated opal [12], we could observe the influence of the interface between the substrate and the compact opal on the reflectivity. In this section, we report on dedicated experiments performed on an opal with much smaller spheres and white light irradiation, in order to test some of our predictions related to the Bragg-type reflectivity peak.

### 1) **Description of the experiment**

Our sample was made of 20 layers of $D = 280$ nm (+/- 5%) microspheres of $S_iO_2$. The Langmuir-Blodgett deposition technique was used to fabricate this opal on a microscope slide (glass). Such a substrate has an index close to the one of the $SiO_2$ microspheres, and the parallelism of such a substrate is poor enough to make internal interferences negligible.

The sample was illuminated with a collimated white light from a fibered type supercontinuum source (LEUKOS SM 20, for 400-1700 nm, pulses < 1ns, repetition rate: 20Hz, average power > 40 mW). The beam diameter was around 2 mm, corresponding to an averaging on many spheres (at this scale, each layer of the opal is already polycrystalline).

The beams of interest (the reflected one, or the transmitted one) are collected by a second optical fiber connected to a fibered spectrometer (Ocean Optics USB 2000+, detection for $\lambda$ = 200-1100 nm, resolution $\Delta\lambda \sim$ 0.4 nm). The orientation of this detection fiber must be carefully aligned to optimize the detected level and to ensure that the collection efficiency is the same all over the spectrum. This is an important source of uncertainty for a multimode fiber and for an analysis over a rather broad spectra range, imposing some limits to our experimental accuracy. Reflection and transmission spectra of the opal are obtained once normalized against the spectral content of the white source, whose temporal fluctuations are notable so that frequent analyses of its spectrum are needed. For sensitivity reasons, reliable spectra are obtained mostly in the 400-800 nm region, covering here a 1.5 - 3 range of $\lambda/D$ values.

The reflection and transmission spectra were recorded for various incident angles θ, both for an irradiation on the vacuum side or on the substrate side. The incident polarization was controlled by an external polarizer. Our experiments confirm that the polarization of the beam after reflection or transmission in the opal is mostly conserved [18].

### 2) <u>Comparison between experimental and theoretical results</u>

The introduction of an *ad hoc* absorption coefficient is essential for the model, in order to avoid the unphysical prediction R + T = 1 for a transparent medium where scattering is ignored. In the previous section III, we have applied the formalism of section II to evaluate the opal reflectivity, rather than transmission, but transmission is derived by the same formal calculation developed in section II. Hence, we can evaluate the absorption coefficient needed in the model by a comparison with the experimental transmission spectra (see Fig. 9). Experiments clearly show a wavelength dependence in the transmission, which drops down in the short wavelength region, although glass remains transparent. We have investigated various power-laws for the wavelength dependence of the absorption coefficient, and have found a reasonable agreement for the transmission spectrum for an absorption coefficient resembling a Rayleigh-like scattering, $\alpha(\lambda) = \frac{\beta}{\lambda^4}$ with $\beta$ a constant. In all the following, we take $\beta$ = 1.5 $10^{-20}$ m$^3$. The hole observed in the transmission spectrum around 550 nm is nothing else than a reduced transmission associated to the strong Bragg reflection. Remarkably, its shape, notably its amplitude and width, which is expected to depend on the choice of the absorption coefficient, is satisfactorily reproduced by our calculation (fig 9 a, b). This agreement persists when changing the side of irradiation for this transmission measurement (*i.e.* propagation through the substrate and the opal, or propagation through the opal and the substrate), or the polarization. When increasing the incidence angle (fig 9 c, d), there is still a reasonable agreement between the experiment and the prediction, although we have not adjusted any parameter in the model, keeping the same amount of absorption - per unit of length "traveled" by the light in the stratified model -.

In figure 10, we have represented the reflectivity spectra for a large range of incidence angles, comparing TE and TM polarizations, and irradiation on the opal and substrate sides. We have also plotted the corresponding predictions, for a model in which the same set of parameters is used for all these various experiments. The choice of the explored $\lambda/D$ range implies the presence of a Bragg reflection peak, whose exact position should vary with the incidence angle. We observe indeed a very good agreement between the experimental and theoretical positions of these peaks, whose wavelength becomes smaller when increasing the incidence angle. In fig 11, we have plotted the experimental peak wavelength for reflection as a function of the incidence angle. It is compared on the one hand with the predictions of our model, which includes both the periodical regions of the opal and the peripheral half layers, and on the other hand with the simplified Bragg model taking into account a global effective index. Although the differences remain marginal, our model seems more precise than the one with a global effective index.

A specific added value of our model is that it predicts the width and amplitude of this Bragg peak, as it already did satisfactorily for transmission. Here, the width estimates always appear in an acceptable agreement with the experiment. Our model is even able to produce sometimes an excellent fitting (*e.g.* 30° and 20 ° in TM) with the experiment, including a satisfactory agreement for the small oscillations, shown to be Fabry-Perot type (see III-4). The quantitative discrepancies in the reflection amplitude may be mostly traced back to experimental defects in the opal fabrication. In particular, when the Bragg peaks appear smaller than predicted, it is natural to consider that the opal, with its successive layer-by-layer deposition, is not as periodical as in the ideal calculation. We also observe a satisfactory trend of the overall reflectivity (away from the Bragg peak) when varying the incidence angle, with differing behaviors for TE and TM polarizations. In particular, in TM polarization (figs.10 b, d), the overall reflectivity decreases close to zero for large angles especially for irradiation from substrate side -*i.e.* large contrast between the input medium (substrate) and the first half-layer "gap"- . This is expected and is analogous to a near Brewster incidence angle, when the reflectivity undergoes a string influence of the first half layer. A more intriguing observation is on Fig. 10d for $\theta = 50°$, where the residual experimental Bragg peak has a higher amplitude than predicted. This could originate in a residual depolarization of the reflected - or "backscattered"- light, or simply in an imperfect polarization of the light reaching opal. A marginal mixture of TE and TM polarizations inside the opal which would hence explain the observation of a Bragg peak when calculations rigorously limited to TM are not able to justify such an effect.

V)     **<u>CONCLUSION AND PERSPECTIVES</u>**

To summarize, our simple model based on a stratified effective index in a direction perpendicular to the opal (*z*) has permitted to evaluate the main features of reflection and transmission for on an opal deposited on a substrate, with the ability to allow for quantitative predictions. A major limitation of such a one-dimensional model is that it cannot include the details of the crystalline structure of the opal with its various symmetry planes [*e.g.* the reflection on a (200) plane]; it is however well-suited to include the partial disorder of a real opal, notably when fabricated by LB methods. Also, the need of an *ad hoc* extinction coefficient to describe scattering is susceptible to vary strongly with the wavelength (to a lesser extent with the incidence angle), and this may limit the range of validity of the quantitative predictions. Our model makes it easy to discriminate between various physical effects. The periodic part of the stratified index is, for example, responsible of the Bragg peak in reflection; it is an originality of our approach that our rather elementary calculation allows realistic predictions the height and width of this peak, whose precise position can be compared with predictions of cruder models, and that it can even predict quantitative values for the successive higher order Bragg peaks. Small Fabry-Perot type oscillations due to the interferences between the first and last layer can also be recognized. However, the low crystalline quality of our sample cannot warranty the exact thickness of the sample, and this does not allow to check experimentally the exact phase of these oscillations. Conversely, our layered approach helps to understand that there is a complex interference system between all the thin layers of the opal. This clearly justifies the wavy behaviors of reflection/transmission spectra, which appear modified when introducing a small change in the internal structure.

The demonstration of the influence of the first half layer, which intrinsically breaks the periodicity by its interface with a substrate of an arbitrary index, or with vacuum, is an important result easily evidenced by our model. The size of the "gap" region, relatively to the wavelength, is an essential parameter. This appears to be a very general situation, although often ignored when one has to evaluate the optics of a photonic crystal. This "interface" coupling, occurring on a wavelength scale can easily hinder internal (in-depth) features, such as defects, of the opal or of a photonic crystal.

The model that we have developed is highly flexible. It is easy to introduce a defective layer in the opal, and it would be also possible to introduce some dispersion in the average layer thickness, or to introduce an extra layer, non compactly arranged as can be imposed by the on-purpose [32] introduction of a special layer. Our one-dimensional model is intrinsically applicable to a compact arrangement of stacked parallel cylinders [22], provided that $f(z)$ is properly redefined in eq. 3. Another natural extension of our approach would be application to the situation of an inverse opal. The formalism may also be applied for an opal deposited on a prism, an interesting situation when the propagation in the interface region (when $\lambda \leq D$) is mostly the propagation of an evanescent wave.

At last, it is possible to consider the situation of an opal infiltrated by some material (liquid, gas, dopants), as if the infiltration is an added defect to the layer structure. In a forthcoming paper, we apply such a model for an infiltration of a resonant material, in order to calculate the resonant optical response of a material infiltrated in a photonic crystal, hence demonstrating that rather remote regions of the opal can contribute to the reflectivity for well-chosen incidences.

**Acknowledgment**

*Work supported by the ANR project "Mesoscopic gas" 08-BLAN-0031. The opal was deposited at CRPP-Bordeaux in the Serge Ravaine group. We acknowledge discussions with Agnès Maitre group.*

**Figure captions**

Fig 1: (a) Solid line: the filling factor $f(z)$ in the $z$-height plane for an opal made of $N$ layers. The dashed lines represent the individual filling factors for the first and second layers; (b) The one-dimensional effective index $n_{eff}(z)$ of the opal, for $n_{sphere} = 1.4$, shown here in discrete version (stratified medium with finite thickness layer).

Fig. 2: Schematic of light propagation in the stratified description. $E_{i,i+1}$ is the tangential component of the electric field at the interface between the $i^{th}$ layer and the $i+1^{th}$ layer. The direction of the plane waves (forward and backward) propagating in the $i^{th}$ layer is governed by the angle $\sin \theta_i = (n_0/n_i) \sin \theta_0$.

Fig 3: Reflection spectrum calculated for $N = 20$ layers. Irradiation on the vacuum side, TE polarization, $\theta = 20°$, $n_{substrate} = 1.4$ ($= n_{sphere}$). The absorption coefficient is: (a) $\alpha D = 0$; (b) $\alpha D = 0.1$.

Fig 4: Reflectivity for the "fused opal" model: (a) effective index of the "fused opal"; (b, c, d, e): wavelength dependence of the reflectivity for the indicated external angles $\theta$ of incidence. The irradiation is (b, c) from the vacuum side or (d, e) from the substrate side. Polarization is TE for (b, d) and TM for (c, e). In (d) and (e), the dotted or dashed lines (red on line) for the smaller values of $\lambda/D$ indicate the Fresnel reflection coefficient for a planar substrate/vacuum interface. Calculations for $\alpha D = 0.1$ and $n_{substrate} = 1.6$.

Fig 5: Wavelength dependence of the reflectivity for the "fused opal" model, for irradiation from the substrate side and under normal incidence. Values of $n_{substrate}$ as indicated. Calculations are for $\alpha D = 0.1$.

Fig 6: Reflectivity spectrum (vacuum side) for a spatially periodic distribution of the effective index $n_{eff}(z)$ as shown in the inset. Calculations are for $N=20$, $\theta = 20°$, $n_{substrate}=1.4$, $\alpha D=0.193$, and TM polarization.

Fig 7: Reflectivity spectra (vacuum side) for different numbers of layers $N_{layer}$ (as indicated). TE polarization, $n_{substrate}=1.4$, $\alpha D = 0.0552$, $\theta = 20°$. The curve for $N_{layer} = 30$ is not visible as just superimposed to the one for $N_{layer} = 10$.

Fig 8: Reflectivity spectra (vacuum side) under normal incidence for various positions of the $i^{th}$ layer carrying a defect (see inset); $n_{substrate}=1.4$, $N=20$, $\alpha D=0.09$.

Fig. 9: Transmission spectra (TE polarization) on a sample with 20 layers of $D = 276$ nm glass spheres: (a, b) vacuum side; (c, d) substrate side, for (a, c) $\theta = 20°$; for (b, d) $\theta = 50°$. Experimental curves (in black) are compared to calculated values (red on line) for which one has taken $\alpha = \beta/\lambda^4$ with $\beta = 1.5 \cdot 10^{-20}$ m$^3$.

Fig. 10: Reflectivity spectra for various incidence angles $\theta$ (as indicated) for the same experimental sample (black) and related calculations (red on line) as in fig. 9: (a, b) on vacuum side; (c, d) on substrate side; with polarization (a, c) TE; (b, d): TM.

Fig. 11: Position of the peak of the reflectivity spectrum ($R_{peak}$) as a function of the incidence angle (substrate side). Comparison between the experiment, our calculations, and the simplified Bragg equation (sample and calculations are the same as in fig.10, the polarization is TE).

FIGURE 1

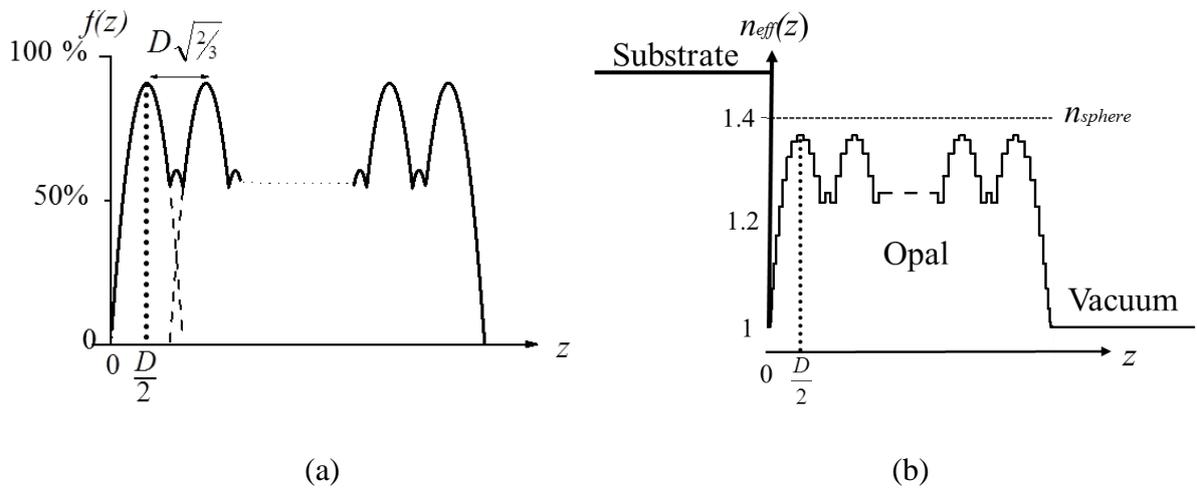

(a)  (b)

FIGURE 2

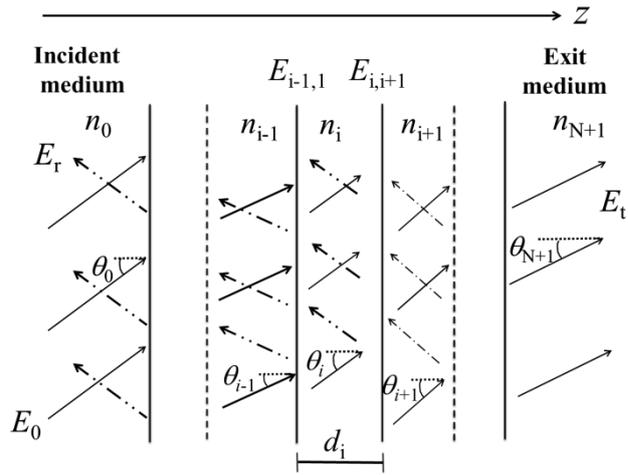

FIGURE 3

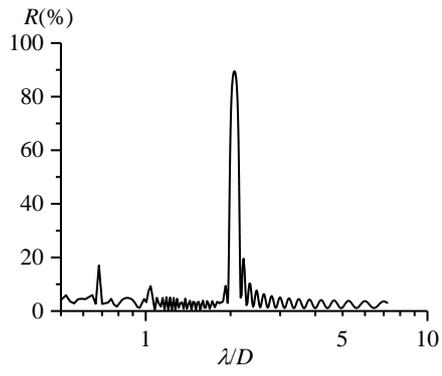
(a)

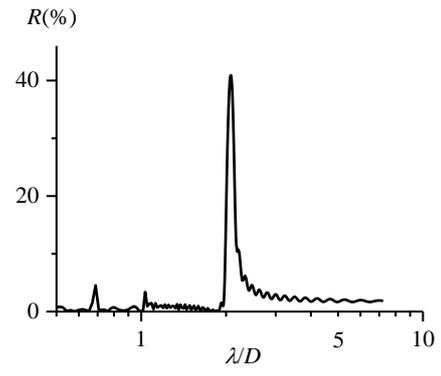
(b)

FIGURE 4

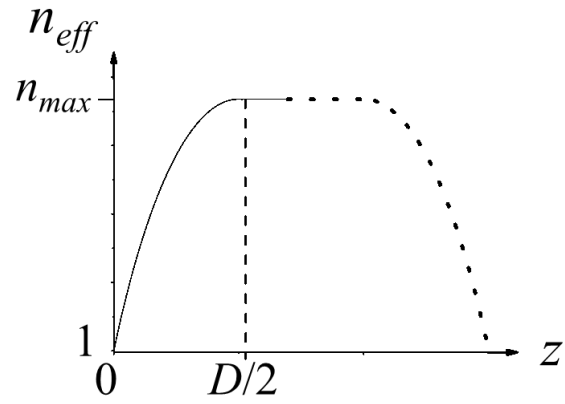

(a)

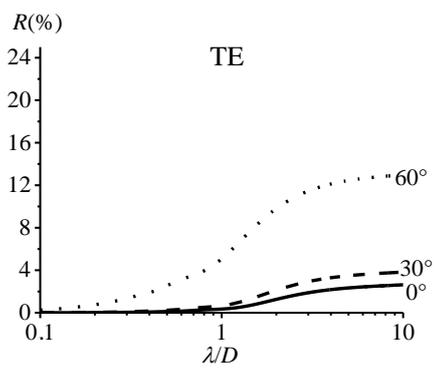

(b)

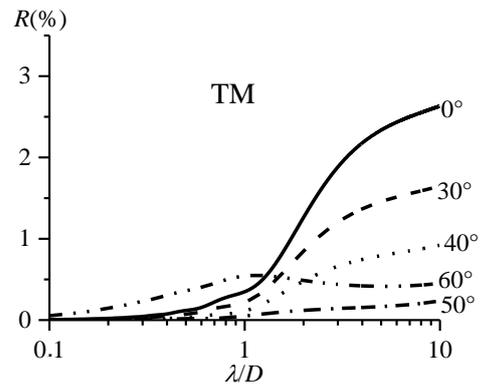

(c)

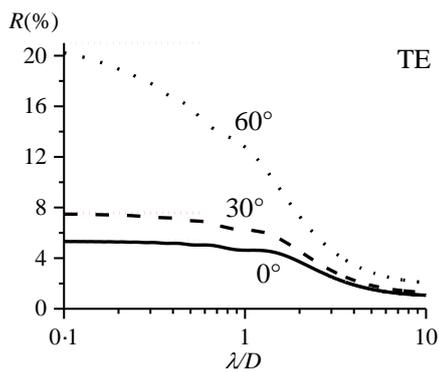

(d)

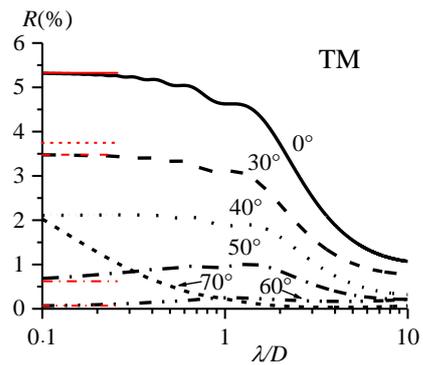

(e)

FIGURE 5

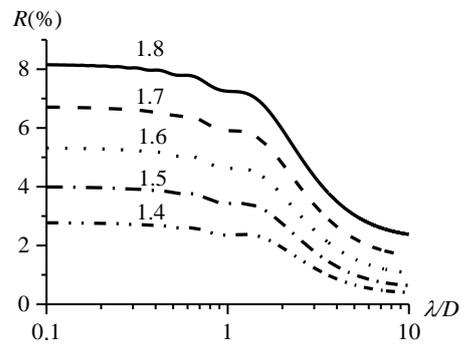

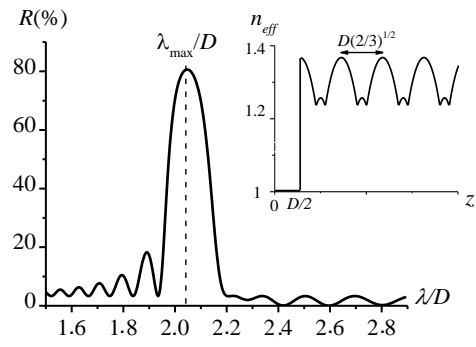

FIGURE 7

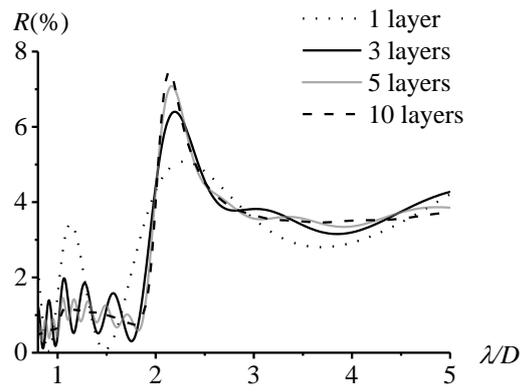

FIGURE 8

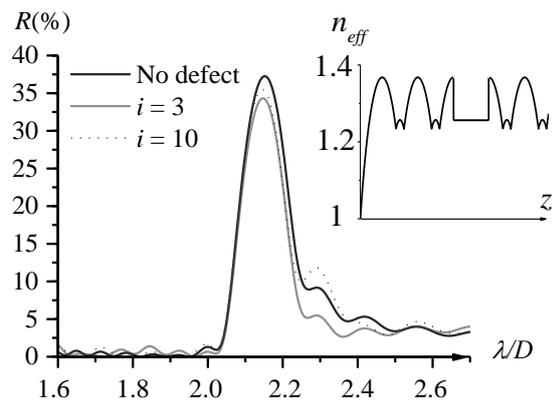

FIGURE 9

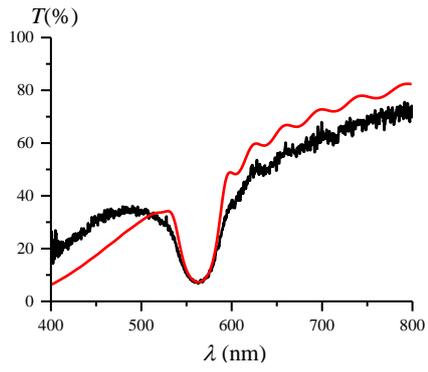

(a)

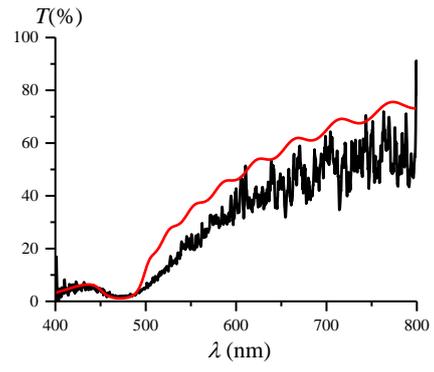

(b)

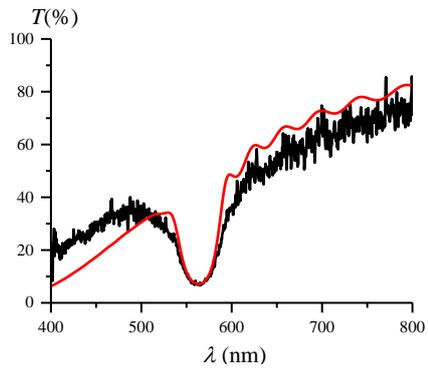

(c)

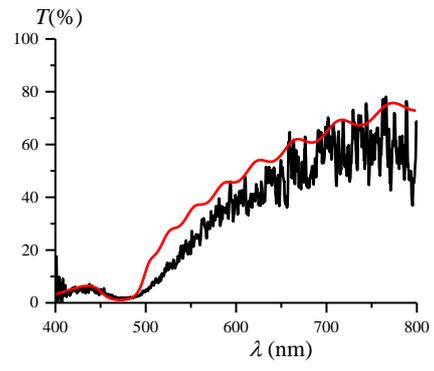

(d)

FIGURE 10

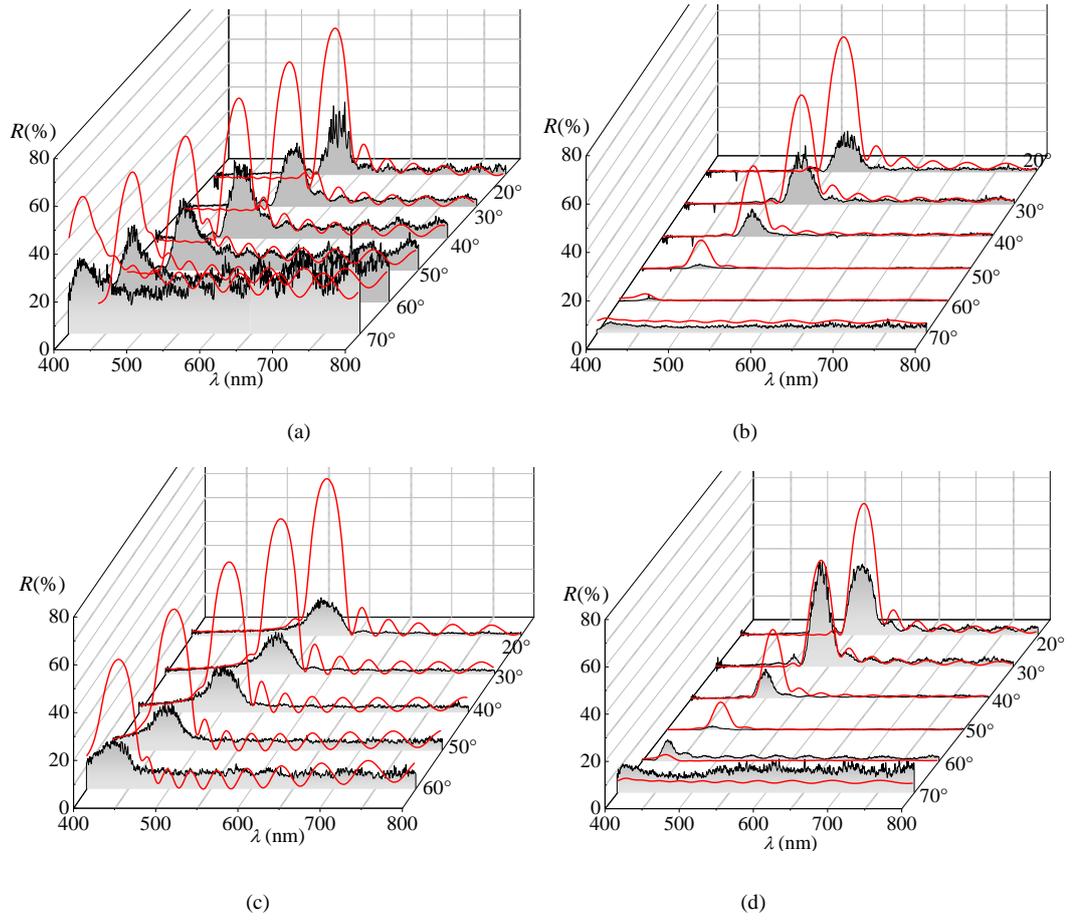

(a)

(b)

(c)

(d)

FIGURE 11

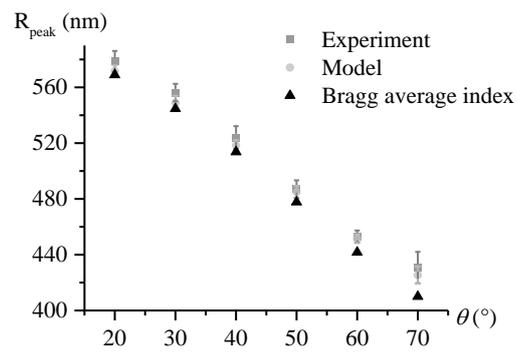